\documentclass[11pt,twoside,a4paper,cr,final]{cms-tdr}
\def\svnVersion{65710:67246MP}

\begin{document}

\hyphenation{had-ron-i-za-tion}
\hyphenation{cal-or-i-me-ter}
\hyphenation{de-vices}

\RCS$Revision: 62698 $
\RCS$HeadURL: svn+ssh://svn.cern.ch/reps/tdr2/papers/XXX-08-000/trunk/XXX-08-000.tex $
\RCS$Id: XXX-08-000.tex 62698 2011-06-21 00:28:58Z alverson $

\newcommand{\xt} {\ensuremath{x_{\mathrm{T}}}~}

\newcommand{\microbinv} {\mbox{\ensuremath{\,\mu\text{b}^\text{$-$1}}}~}

\newcommand{\RAA} {\ensuremath{R_{\mathrm{AA}}}~}
\newcommand{\TAA} {\ensuremath{T_{\mathrm{AA}}}~}
\newcommand{\RCP} {\ensuremath{R_{\mathrm{CP}}}~}
\newcommand{\sNN} {\ensuremath{\sqrt{s_{_{\mathrm{NN}}}}}~}
\newcommand{\Npart} {\ensuremath{N_{\mathrm{part}}}~}
\newcommand{\Ncoll} {\ensuremath{N_{\mathrm{coll}}}~}

\newcommand{\HYDJET} {\textsc{hydjet}~}
\newcommand{\PYQUEN} {\textsc{pyquen}~}
\newcommand{\AMPT} {\textsc{ampt}~}

\cmsNoteHeader{2011/108} % This is over-written in the CMS environment: useful as preprint no. for export versions
\title{Nuclear modification factors from the CMS experiment}

\author{Yen-Jie Lee for the CMS collaboration
}
\address{
\vspace{2mm}
{\scriptsize
Massachusetts Institute of Technology, 77 Mass Ave, Cambridge, MA 02139-4307, USA\\
}}

\date{\today}

\abstract{
We report the measurements of nuclear modification factors of the Z bosons, isolated photons and charged particles in $\sqrt{s_{NN}}$ = 2.76 TeV PbPb collisions with the CMS detector. The nuclear modification factors are constructed by dividing the PbPb $p_T$ spectra, normalized to the number of binary collisions, by the pp references. No modifications are observed in isolated photon and Z boson production with respected to the pp references while large suppression is observed in the charged particles. 
\vspace{8mm}

\begin{center}
Presented at \textit{QM2011}: \textit{Quark Matter 2011}
\end{center}

}

\hypersetup{%
pdfauthor={CMS Collaboration},%
pdftitle={CMS Paper Template 2006 LaTeX/PdfLaTeX version},%
pdfsubject={CMS},%
pdfkeywords={CMS, physics, software, computing}}

\maketitle %maketitle comes after all the front information has been supplied

\section{Introduction}

The hot and dense matter produced in heavy-ion collisions, often referred to as the quark-gluon plasma (QGP), can be studied in various ways. The first way is to measure the yields of colorless particles in heavy-ion (AA) collisions that are unmodified in order to understand the initial state of the collisions. The second approach is to compare measurements made in AA collisions to those in proton-proton (pp) and proton- \mbox{(or deuteron-)} nucleus collisions. At the Relativistic Heavy Ion Collider (RHIC) direct photons play the reference role~\cite{Adler:2005ig}. At the Large Hadron Collider (LHC) energies, not only photons but also Z bosons, decaying to two leptons, become available~\cite{Kartvelishvili:1995fr,ConesadelValle:2009vp}. Since, once produced, photons and leptons from Z decays traverse the produced hot and dense medium without interacting strongly, they provide a direct test of perturbative QCD (pQCD) and the nuclear parton densities~\cite{Arleo:2011gc}. High transverse energy ($E_T$) prompt photons in nucleus-nucleus collisions are produced directly from the hard scattering of two partons. In order to suppress the much larger background coming from the electromagnetic decays of neutral mesons (mostly $\pi^0,\eta\to\gamma\,\gamma$) produced in the fragmentation of the hard scattered partons, isolation requirements on the reconstructed photon candidates are imposed. The application of isolation requirements suppresses also the fragmentation photon component and enhances significantly photons coming from the quark-gluon Compton process~\cite{Ichou:2010wc}.

The measurement of charged particle \PT spectrum is motivated by lower energy results~\cite{Adams:2005dq,Adcox:2004mh,Arsene:2004fa,Back:2004je} 
from the Relativistic Heavy Ion Collider (RHIC), where high-\PT particle production was found to be strongly suppressed relative to expectations from an independent superposition of nucleon-nucleon collisions. This observation is typically expressed in terms of 
the nuclear modification factor:
\begin{equation}
\RAA(\PT) = \frac{d^{2}N^{\mathrm{AA}}/d\PT d\eta}{\left<\TAA\right> d^{2}\sigma^{\mathrm{NN}}/d\PT d\eta},
\label{eqn:def_raa}
\end{equation}
where $N^{\mathrm{AA}}$ and $\sigma^{\mathrm{NN}}$ represent the yield in nucleus-nucleus collisions
and the cross section in nucleon-nucleon collisions, respectively.  
The pseudorapidity is defined as $\eta=-\ln[\tan(\theta/2)]$, with $\theta$
being the polar angle of the charged particle with respect to the counterclockwise beam direction.
The nuclear overlap function $\left<\TAA\right>$ 
is the ratio of the number of binary nucleon-nucleon collisions $\left<\Ncoll\right>$ calculated from a Glauber model 
of the nuclear collision geometry~\cite{glauber}
and the inelastic nucleon-nucleon cross section $\sigma^{\mathrm{NN}}_{\mathrm{inel}}=(64 \pm 5)$\,mb 
at $\sqrt{s}=2.76$ TeV~\cite{PDBook}.  
In the absence of nuclear effects on the PbPb \PT spectrum, the factor \RAA is unity by construction.  
In particular, together with recent studies of jet quenching and fragmentation properties~\cite{jetquenchingCMS,jetquenchingATLAS}, the modification of the \PT spectrum compared to nucleon-nucleon collisions at the same energy can shed light on the detailed mechanism by which hard partons lose energy traversing the medium~\cite{dEnterria:2009am}.

This report presents the measurements of nuclear modification factors of the isolated photons, Z bosons and charged particles as a function of event centrality for PbPb collisions collected by the Compact Muon Solenoid (CMS) experiment at a center-of-mass energy of 2.76 TeV per nucleon pair.

\section{Experimental methods}

Final-state particles in the selected collision events are reconstructed 
in the CMS detector. A detailed description of CMS can be 
found elsewhere~\cite{JINST}. 
A minimum-bias (MinBias) event sample is
collected using coincidences between trigger signals from the $+z$ and $-z$ sides of either the Beam Scintillator Counters (BSC) or the Hadronic Forward Calorimeter (HF, covering $2.9<|\eta|<5.2$) . The minimum bias trigger and event selection efficiency is $97\pm 3\%$.~\cite{Chatrchyan:2011sx}.

The events are also selected by the two-level trigger of CMS. The di-muon trigger requires two muon candidates in the muon detectors at the first hardware level (L1). At the software-based higher-level (HLT), two reconstructed tracks in the muon detectors are required, each with a $p_{\rm T}$ of at least 3~GeV/$c$. The photon trigger requires a L1 electromagnetic cluster with $E_T> 5$ GeV and a HLT photon with $E_T > 15$ GeV. In order to extend the statistical reach of the charged particle \PT spectra, data recorded by single-jet triggers
with uncorrected transverse energy thresholds of $\ET = 35$ GeV and 50 GeV are included in the analysis. The jet-energy thresholds in the trigger are applied after subtracting the underlying event energy but without correcting for calorimeter response. In order to select a pure sample of inelastic hadronic collisions for analysis, the contamination from ultra-peripheral collisions and non-collision beam background are removed following the prescription of Ref.~\cite{Chatrchyan:2011sx}.

\section{Centrality determination}
\label{section:centrality}
\vspace{-0.4cm}

For analysis of PbPb events it is important to determine the overlap or 
impact parameter of the two colliding nuclei, usually called 
``centrality''. Centrality is
determined with the MinBias sample using the total sum of energy signals from both HF, where
the events that deposit the top 10\% most energy of the total interaction cross section are
called the 0-10\% most central (small impact paramter). Using Glauber model simulations, the intervals can be correlated with more meaningful physics quantities. Details of the centrality determination and the model calculations are described in Ref.~\cite{Chatrchyan:2011sx}. The two most commonly used centrality parameters are N$_{\rm part}$, the total number of nucleons in the two Pb nuclei (with mass number 208) which experienced at least one collision, and N$_{\rm coll}$, the total 
number of nucleon-nucleon collisions.

\section {Results}
\subsection {Z boson}

\begin{figure}[hbtp]
  \begin{center}
\includegraphics[width=0.32\textwidth]{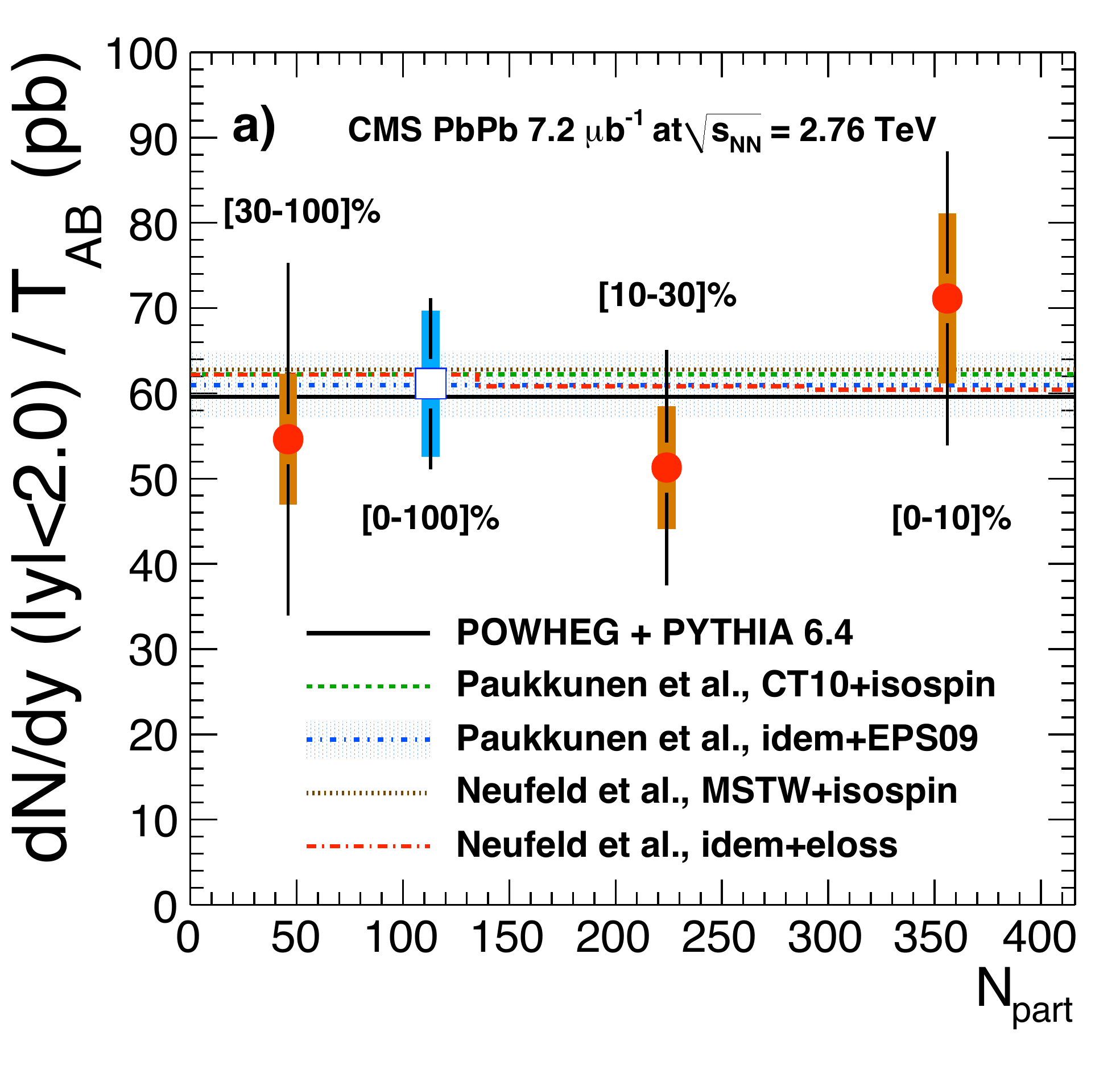}
    \includegraphics[width=0.32\textwidth]{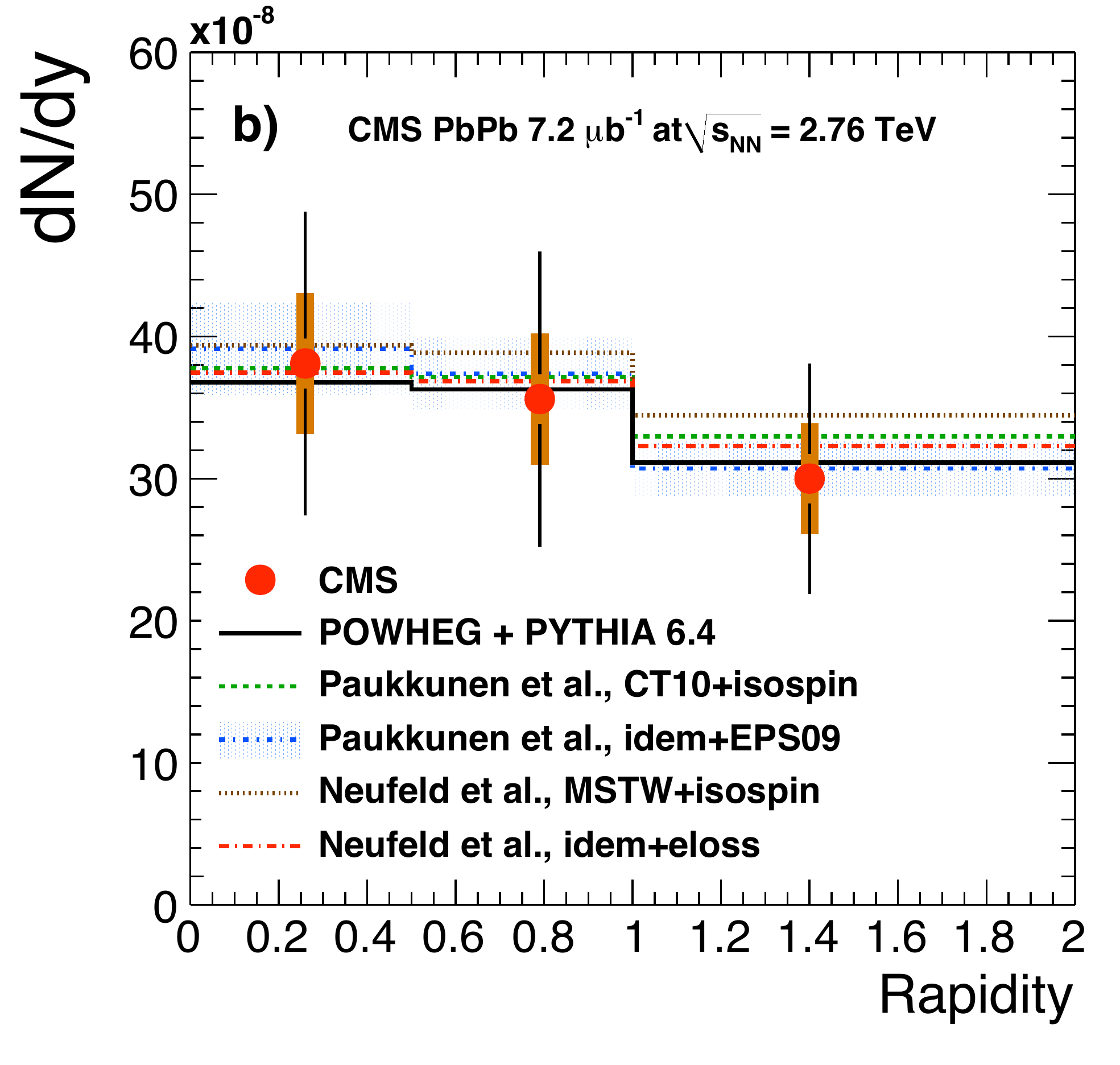} 
    \includegraphics[width=0.32\textwidth]{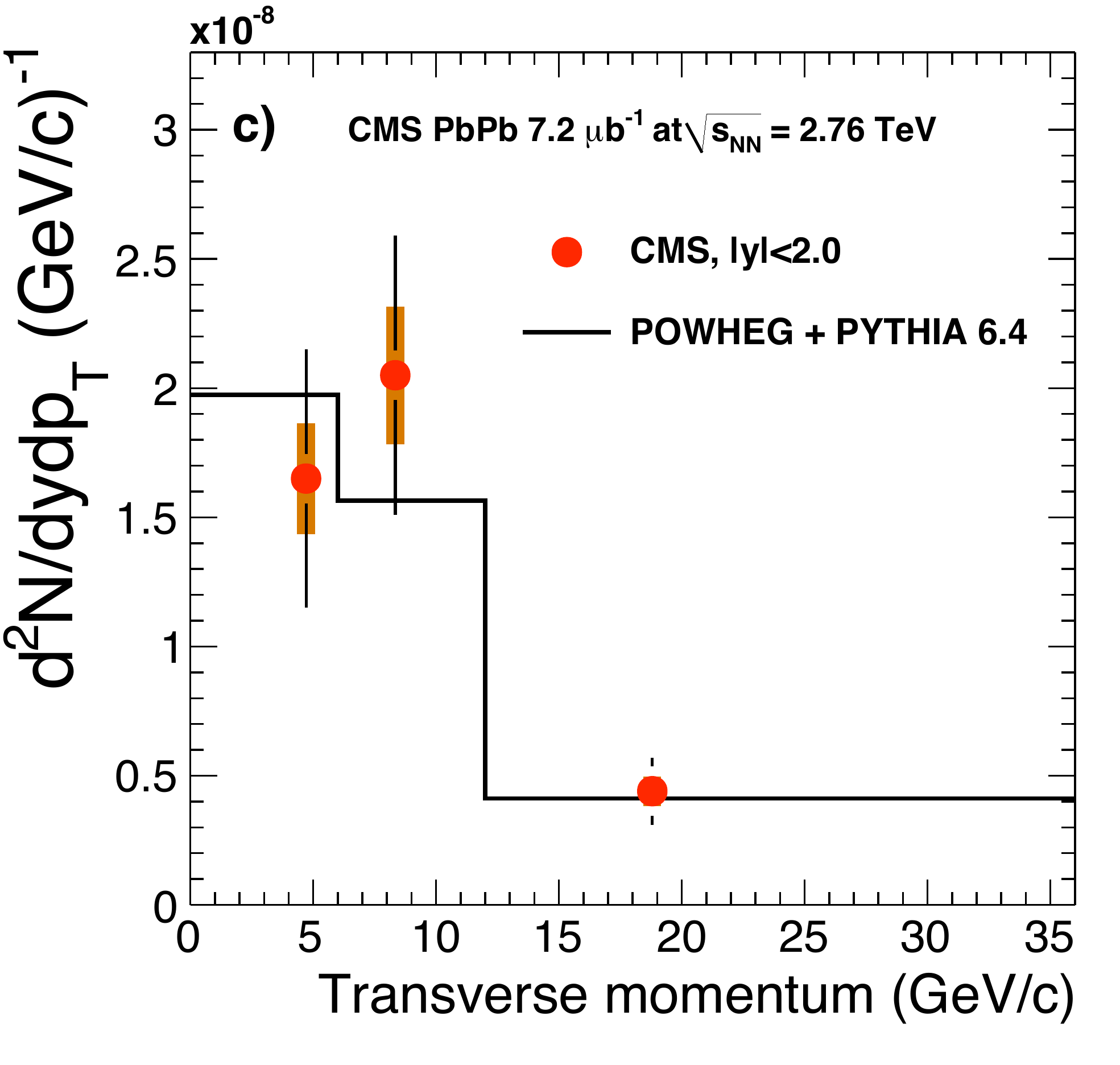}
    \caption{The yields of $Z \rightarrow \mu\mu$ per event: a) $dN/dy$ divided by the expected nuclear overlap function $T_{\rm AB}$ and as a function of event centrality parameterized as the number of participating nucleons $N_{\rm part}$, b) $dN/dy$ versus the Z boson $y$, c) $d^2N/dydp_{\rm T}$ versus the Z boson $p_{\rm T}$. Data points are located horizontally at average values measured within a given bin. Vertical lines (bands) correspond to statistical (systematic) uncertainties. Theoretical predictions are computed within the same bins as the data and are described in the text.}
    \label{fig:yields}
  \end{center}
  \vspace{-2ex}
\end{figure}

The yield of $Z \rightarrow \mu^+ \mu^-$ decays per MB event is defined as $dN/dy (|y|<2.0) = N_{Z} / (\alpha \varepsilon N_{\rm MB} \Delta y )$, where $N_{Z}$ is the number of dimuons counted in the mass window of 60--120~GeV/$c^2$, $N_{\rm MB} = 55 \times 10^{6}$ is the number of corresponding MB events, corrected for trigger efficiency, $\alpha$ and $\varepsilon$ are the acceptance and overall efficiency, and $\Delta y = 4.0$ is the rapidity bin width. The full circles in Fig.~\ref{fig:yields} (a) show the centrality dependence of the Z yield divided by $T_{\rm AB}$, while the open square is for MB events. The variable used on the abscissa is the average number of participating nucleons $N_{\rm part}$ corresponding to the selected centrality intervals, computed in the same Glauber model. No centrality dependence of the binary-scaled Z yields is observed in data~\cite{ZBoson}. A similar result was recently published by the ATLAS collaboration~\cite{Atlas:2010px}.

The normalized yields $(dN/dy)/T_{\rm AB}$ are compared to various calculations~\cite{Eskola:2009uj,Paukkunen:2010qg,Martin:2009iq,Neufeld:2010fj,Alioli:2008gx,Pumplin:2002vw}. Only a marginal centrality dependence is predicted: the inhomogeneous (i.e.~depending on the radial position in nuclei) shadowing is predicted to have negligible impact~\cite{Vogt:2000hp} and the energy-loss prediction drops by 3\% from peripheral to central collisions~\cite{Neufeld:2010fj}. Figures~\ref{fig:yields} (b) and (c) show the differential yields, $dN/dy$ and $d^2N/dydp_{\rm T}$, as a function of the Z boson $y$ and $p_{\rm T}$. The differential yields are compared to the same theoretical calculations as used for the centrality distribution (when available) multiplied by the minimum bias $T_{\rm AB}$ value. In all bins, no significant deviations from binary-collision scaling are observed.

Nuclear modification factors, $R_{\rm AA}=dN / (T_{\rm AB} \times d\sigma_{pp})$, are computed from the AA measured yields $dN$, the nuclear overlap function $T_{\rm AB}$, and the pp$\rightarrow$~Z cross sections $d\sigma_{pp}$ given by the \textsc{powheg} calculation (solid lines on Fig.~\ref{fig:yields}, e.g. $d\sigma_{pp}/dy = 59.6$~pb in $|y|<2.0$). It is found that the $R_{AA}= 1.20 \pm 0.29 {(\rm stat.)} \pm 0.16 {(\rm syst.)}$ in 0-10\% central PbPb collisions which is consistent with unity.

\subsection {Isolated photons}

The corrected photon yield with $|\eta|<1.44$ per MinBias event is defined as $dN/dE_{T} = N^{\gamma}/(\epsilon \times N_{\rm MB} \times \Delta E_{T})$, where $N_{\rm {MB}}=5.19\times 10^{7}$ is the number of MinBias events corresponding to the number of events recorded in the photon-triggered sample and $\epsilon$ is the efficiency of the photon identification. The systematic uncertainty of the photon yield $dN/dE_{T}$ is dominated by the uncertainty of the background estimation which is in the level of 11-31\% and the total systematic uncertainties are  21-37\%.

\begin{figure}[htb]
\begin{center}
\resizebox{0.45\textwidth}{!}{\includegraphics{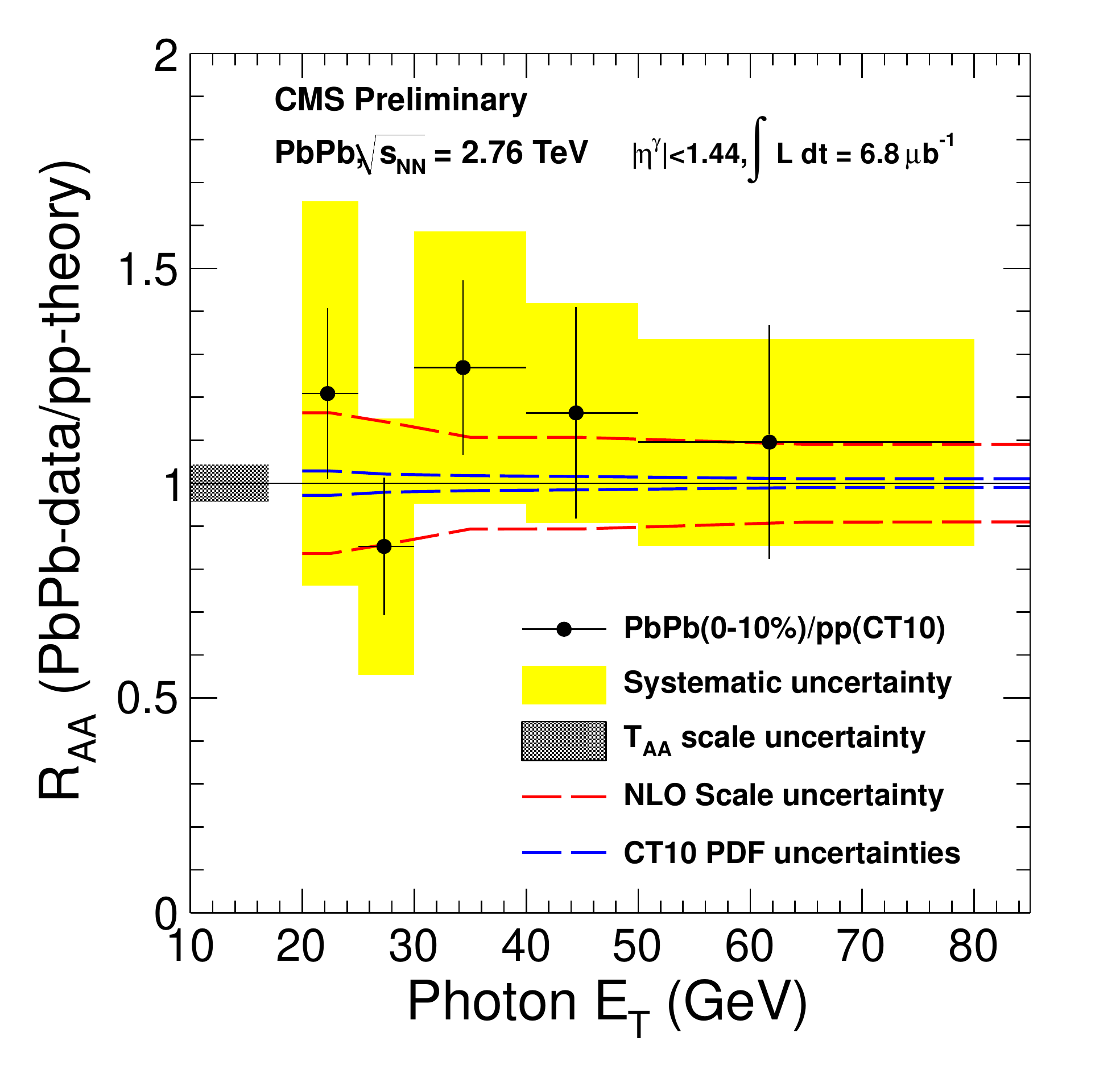}}
\resizebox{0.45\textwidth}{!}{\includegraphics{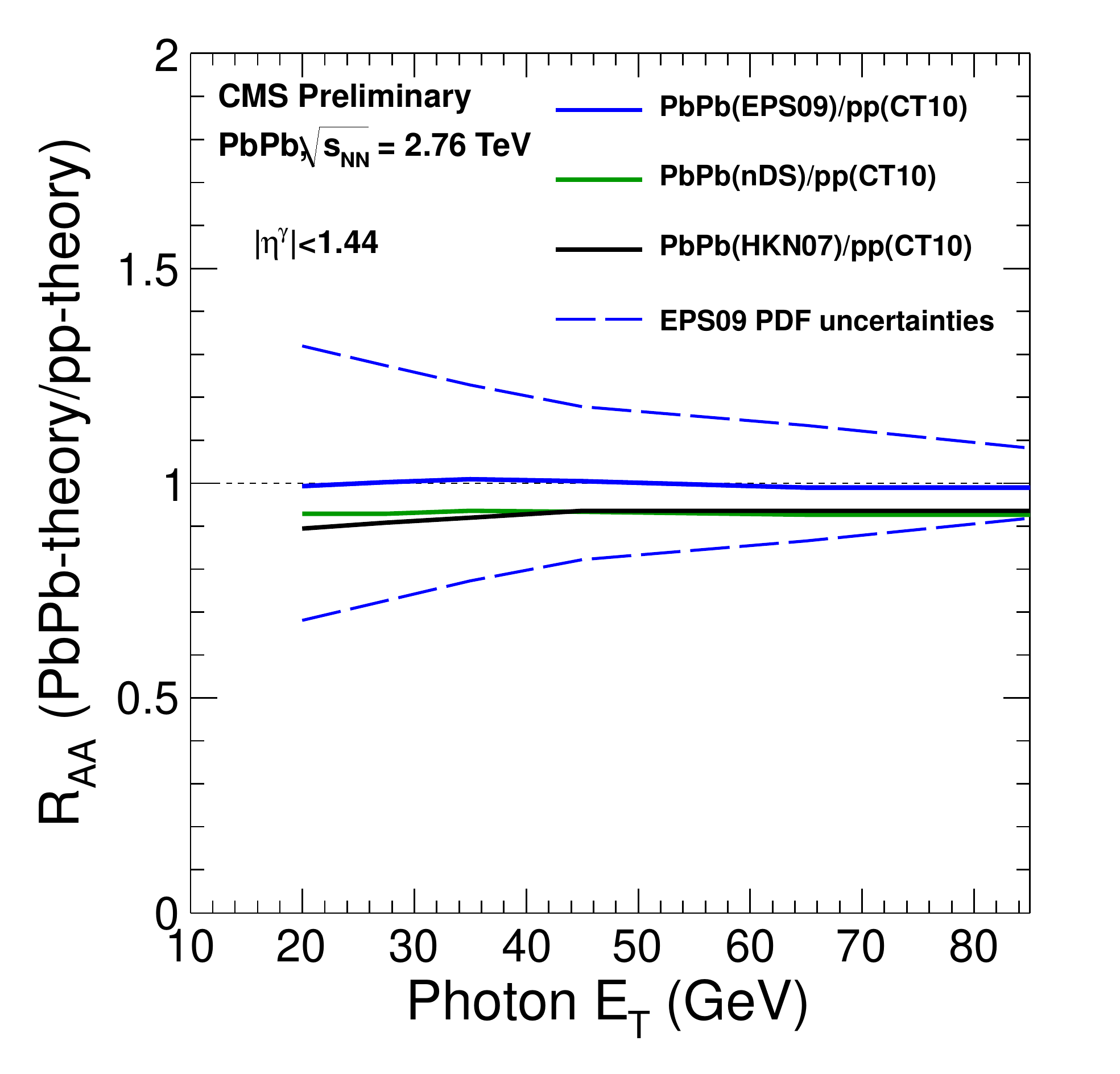}}
\caption{\label{fig:RAA} (Left Panel) Nuclear modification factor $R_{AA}$ as a function of the photon $E_{T}$ measured in 0-10\% central PbPb collisions over the $T_{AA}$-scaled pp JETPHOX prediction at $\sqrt{s_{NN}} = 2.76$~TeV. The vertical lines are the statistical uncertainty. The systematic uncertainties of the measurements, including the uncertainty of $T_{AA}$, are shown as yellow filled areas. The scale uncertainty of the pp calculation is shown as red lines. (Right Panel) The ratio of PbPb and pp JETPHOX predictions as described in the text is shown as the blue line. The uncertainty in this ratio due to the EPS09 PDF is shown as the blue dashed lines. Comparison between the JETPHOX predictions with alternative nPDFs is also shown.}
\end{center}
\end{figure}

The nuclear modification factor $R_{AA}= dN^\gamma/(T_{AA}\times \sigma^\gamma_{pp})$, is computed from the PbPb measured yield $dN^\gamma$, the inclusive isolated photon cross-section $\sigma^\gamma_{pp}$ given by the JETPHOX calculation~\cite{Catani:2002ny}. The uncertainty of $T_{AA}$ is included in the $R_{AA}$ systematic uncertainty. 
Figure~\ref{fig:RAA} (left) shows the $R_{AA}$ as a function of the photon $E_T$ in the 0-10\% central collisions. The results are found to be compatible with unity within the quoted uncertainties. Figure~\ref{fig:RAA} (right) shows the JETPHOX predictions for the same ratio using different nPDFs.

\begin{figure}[htb]
\begin{center}
\resizebox{0.65\textwidth}{!}{\includegraphics{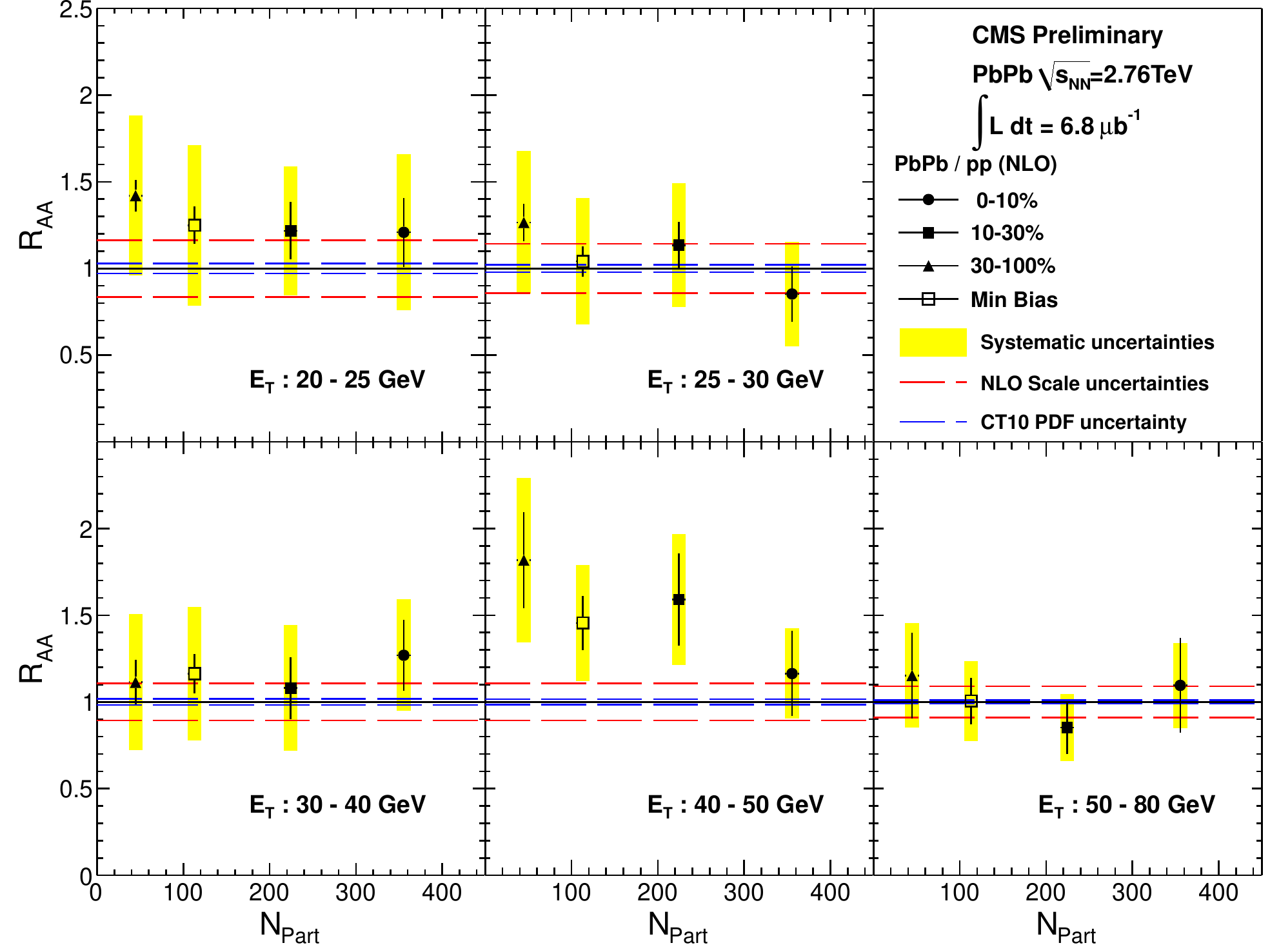}}
\caption{\label{fig:RAAvsNpart} The measured nuclear modification factor $R_{AA} $ as a function of $N_{\rm part}$ for the five different photon transverse energy intervals. The vertical lines are the statistical uncertainty. The systematic uncertainties of the measurements, including the uncertainty of $T_{AA}$, are shown as yellow filled areas. The scale uncertainty of the pp calculation is shown as red lines.}
\end{center}
\end{figure}

Figure ~\ref{fig:RAAvsNpart} shows the $R_{AA}$ as a function of $N_{\rm part}$ in order to investigate the centrality dependence of the nuclear modification effect. No significant centrality dependence is observed in data. 

\subsection{Charged particles}

\begin{figure}[hbtp]
   \begin{center}
        \includegraphics[width=0.65\textwidth]{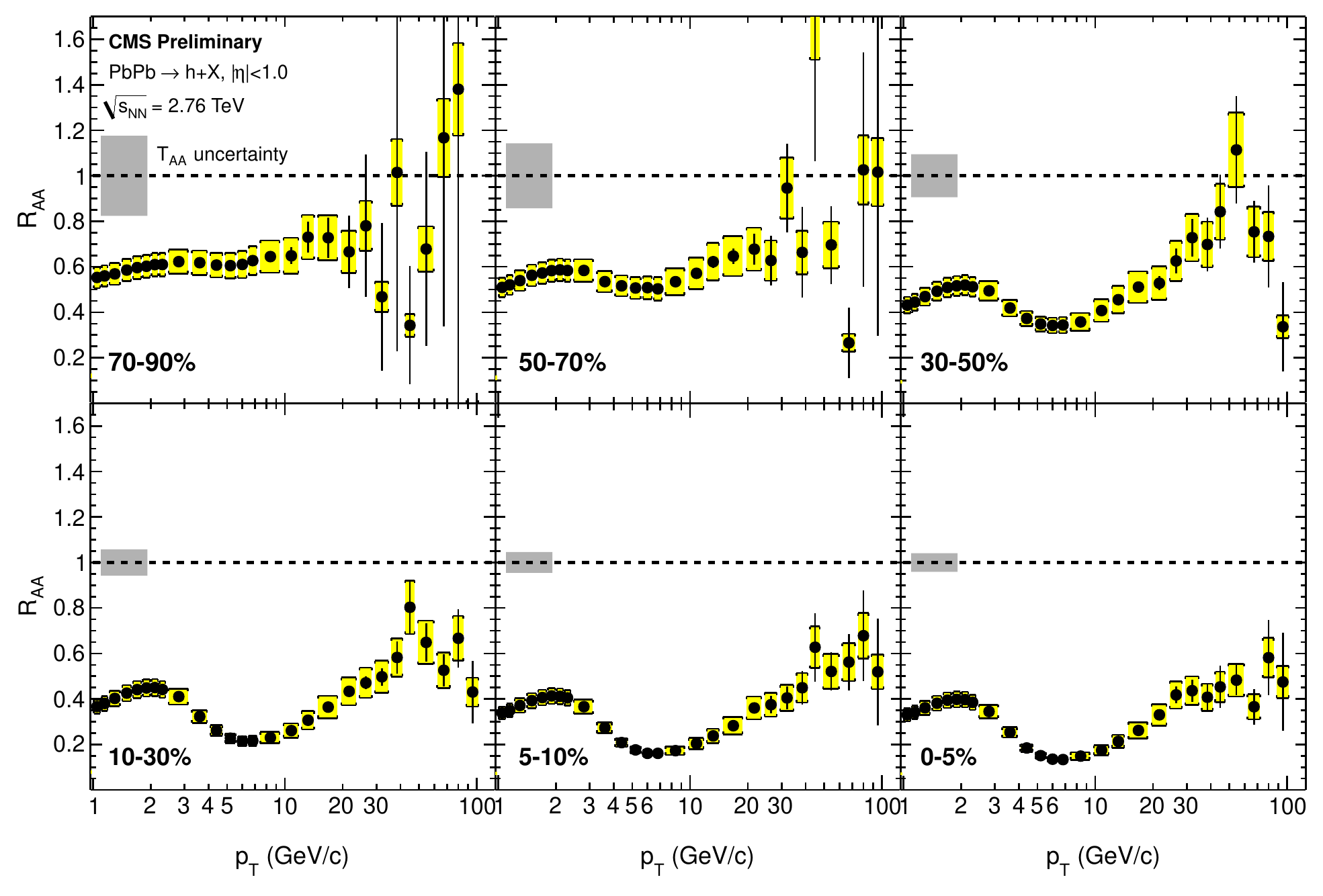} 
	\caption{Nuclear modification factor \RAA (filled circles) as a function of \PT for six centrality intervals. 
	The error bars represent the statistical uncertainties, and the yellow boxes the \PT-dependent systematic uncertainties on the \RAA measurements.  
	An additional systematic uncertainty from the normalization of \TAA, common to all points, is shown as the shaded band around unity
	in each plot.}
     \label{fig:specRaaLogX}
   \end{center}
\end{figure}

\begin{figure}[hbtp]
   \begin{center}
      \includegraphics[width=0.5\textwidth]{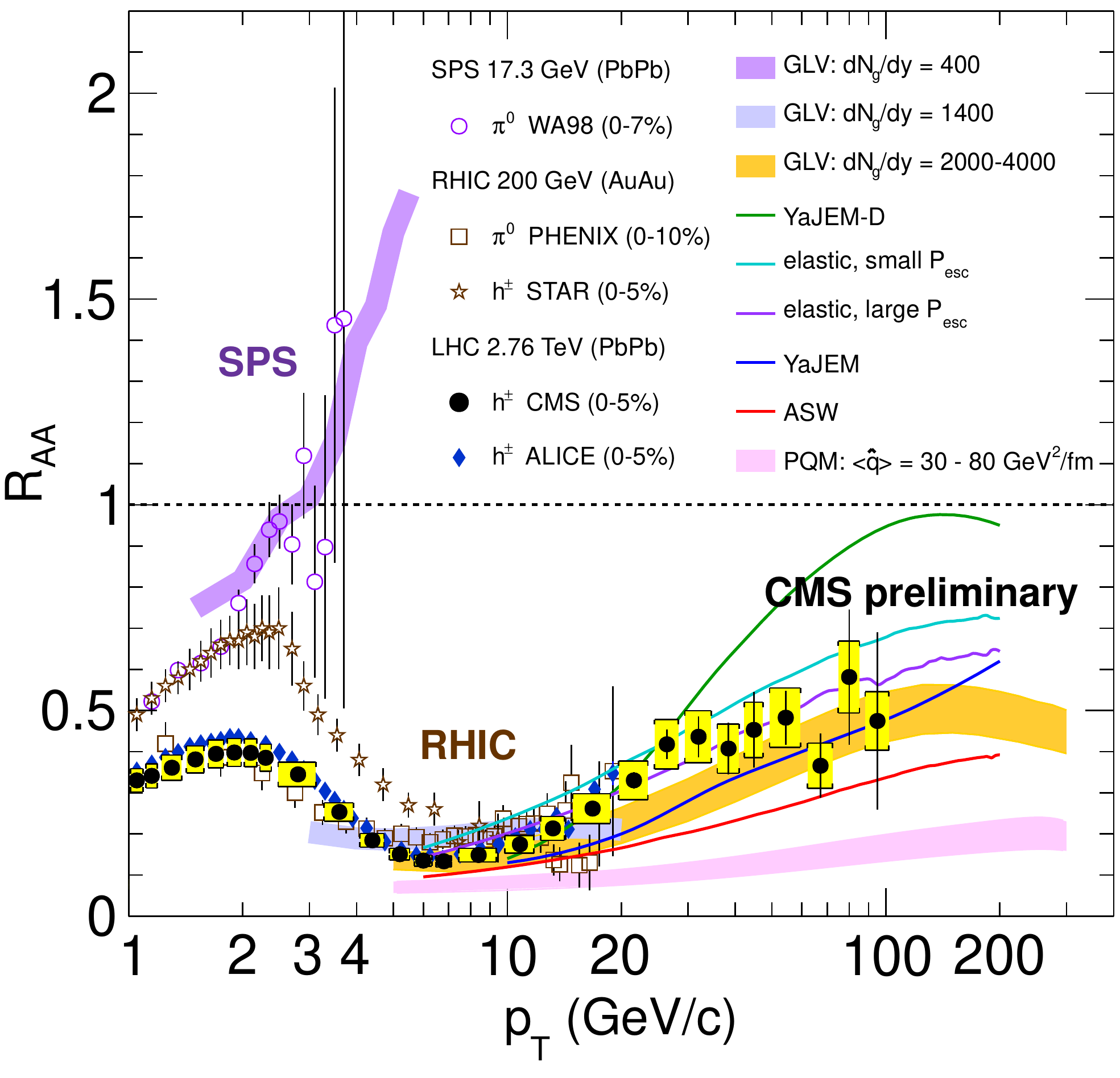}
      \caption{Measurements of the nuclear modification factor \RAA in central heavy ion collisions at three different center-of-mass energies
      as a function of \PT for neutral pions and charged hadrons~\cite{wa98,wa98revisited,phenix,star},
      compared to several theoretical predictions~\cite{pqm,glv,glv2,renk}.  The error bars around the points
      are the statistical uncertainties, and the yellow boxes around the CMS points are the systematic uncertainties.
      The bands for several of the theoretical calculations represent their uncertainties.}
            \label{fig:raaCMSvsTheoryLinY}
   \end{center}
\end{figure}

The nuclear modification factor \RAA is constructed according to Eq.~(\ref{eqn:def_raa}) by dividing the PbPb results by the scaled pp reference obtained from an interpolation~\cite{ppreferenceCMS}.  \RAA is presented as a function of transverse 
momentum in Fig.~\ref{fig:specRaaLogX} for each of the six centrality bins.
In the most-peripheral events (70--90\%), a moderate suppression of about two ($\RAA = 0.5$) is observed at low \PT
with \RAA rising gently with increasing transverse momentum.  
The suppression is increasingly pronounced in more-central collisions, 
as expected from the longer average path lengths traversed by hard-scattered partons as they lose energy via jet quenching.
\RAA reaches a minimum value of 0.13 around 6--7 GeV/$c$ in the 0--5\% bin.  
At higher \PT, the value of \RAA rises and levels off above 40 GeV/$c$ at a
value of approximately 0.5.  A rising \RAA may simply reflect the flattening of the unquenched nucleon-nucleon spectrum at high \PT, 
although the magnitude of the rise is not strongly constrained by theory (see Fig.~\ref{fig:raaCMSvsTheoryLinY}).

The evolution of the nuclear modification factor with center-of-mass energy from SPS~\cite{wa98,wa98revisited} to RHIC~\cite{phenix,star} 
to the LHC~\cite{aliceRAA} is presented in Fig.~\ref{fig:raaCMSvsTheoryLinY}.  
Note that results are shown for both charged hadrons and neutral pions, the latter being somewhat more suppressed
at low \PT (\eg, parton recombination can produce an excess of protons~\cite{starPID}).  At RHIC, the values of \RAA 
from the two different measurements are seen to converge above roughly 8 GeV/$c$~\cite{phenix,star}.
At low \PT, the spectrum of charged particles is significantly more suppressed at the LHC than at RHIC,
although the minimum values and the \PT at which they occur are quite similar.
It remains to be seen whether the similarity between the $\pi^0$ suppression observed by PHENIX at $\sNN=200$GeV~\cite{phenix}
and the charged particle suppression presented in this paper at $\sNN=2.76$TeV is more than a coincidence.  

The CMS measurement of \RAA presented in this paper in the 0--5\% centrality interval can be compared to the published 
ALICE~\cite{aliceRAA} result over the \PT range measured by ALICE.  
Note that the CMS pp reference~\cite{ppreferenceCMS}, derived from an interpolation
that includes CDF and CMS measurements from $\sqrt{s}=0.63$--7TeV, is roughly 5--15\% higher than the ALICE pp reference 
quoted in their paper~\cite{aliceRAA}.  The two results are in agreement within their respective statistical and systematic uncertainties.

The high-\PT measurement of \RAA from this analysis, up to $\PT = 100$ GeV/$c$, is compared to a number of model predictions, both for the LHC design energy of $\sNN=5.5$TeV 
(PQM~\cite{pqm} and GLV~\cite{glv,glv2}) and for the actual 2010 collision energy of $\sNN=2.76$TeV (ASW, YaJEM, and an elastic scattering energy-loss model with parameterized escape probability~\cite{renk}).  While most models predict the generally
rising behavior that is observed in the data at high \PT, the magnitude of the predicted slope varies greatly depending on the details of 
the jet quenching implementation.
The new CMS measurement should help constrain the quenching parameters used in these models and further the 
understanding of the energy-loss mechanism.

\section{Summary}

In conclusion, the Z boson and isolated photon yields in PbPb collisions at $\sqrt{s_{\rm NN}} = 2.76$~TeV have been measured as a function of  centrality. Within uncertainties, no modification is observed with respect to theoretical next-to-leading order pQCD proton-proton cross sections scaled by the number of elementary nucleon-nucleon collisions. This measurement confirms the validity of the Glauber scaling for perturbative cross sections in nucleus-nucleus collisions at the LHC.

A dramatic suppression of the charged particle spectrum has been
observed for the most-central PbPb events, as was previously seen from the lower collision energies at RHIC.  The result is consistent with 
the values of \RAA measured by the ALICE experiment within experimental uncertainties.  
The rise in \RAA from $\PT=6$--40GeV/$c$ is reproduced in the predictions of a number of models, although the magnitude of the rise 
varies substantially between different predictions. Together with  measurements of inclusive jet spectra, fragmentation functions, and energy balance, this measurement of the nuclear modification factor as a function of \PT and collision centrality will help elucidate the mechanism of jet quenching and the properties of the medium produced in heavy ion collisions.

\end{document}